\documentclass{PoS}

\title{Double Hairpin Diagrams and the Planar Equivalence of N=1 Supersymmetric Yang-Mills Theory and One-Flavor QCD}

\ShortTitle{Planar Equivalence of N=1 SUSY Yang-Mills Theory and One-Flavor QCD}

\author{\speaker{Patrick Keith-Hynes}, H.B. Thacker\\
        Department of Physics, University of Virginia\\
        E-mail: \email{hbt8r@virginia.edu}}

\abstract{Recent work by Armoni, Shifman, and Veneziano suggests a large-N equivalence between supersymmetric 
Yang-Mills Theory and one-flavor QCD.
One consequence of this "orientifold projection" is that scalar and pseudoscalar mesons in one-flavor QCD should 
have degenerate mass
since they lie within the same Wess-Zumino supermultiplet.  We use lattice calculations to investigate the mass shifts
caused by "double-hairpin" annihilation diagrams in quenched QCD to test for this degeneracy. Similar
quark-antiquark annihilation processes are studied in the 2-dimensional $CP^{(N-1)}$ model with quenched fermions.
}

\FullConference{XXIVth International Symposium on Lattice Field Theory\\
                July 23-28, 2006\\
                Tucson, Arizona, USA}

\begin{document}

\section{INTRODUCTION}

     The theoretical connections between QCD and string theory which have emerged in recent years have provided a new 
framework for the study of low energy hadron dynamics.  While these connections are, for the most part, only convincingly 
argued in the limit of large $N_c$, it is of great interest to test the predictions of string/gauge correspondences 
against real-world $N_c$ = 3 QCD using both phenomenology and lattice results.  A particularly interesting correspondence 
in this regard is the recently discovered orientifold planar equivalence between {\cal N} = 1 SUSY Yang-Mills theory and ordinary 
nonsupersymmetric one-flavor QCD \cite{ASV1,ASV2,ASV3}.  Although this is a direct field theory to field theory 
equivalence, it has its roots in string/gauge duality.  To the extent that this equivalence is reliable at $N_c$ = 3, 
it provides a powerful new source of insight into low energy quark dynamics.  What is truly remarkable about this 
connection is that the quark of the daughter theory (one-flavor QCD) is the orientifold projection of a gluino in the 
parent theory ({\cal N} = 1 SUSY YM).  If this equivalence is even qualitatively valid at $N_c$ = 3, it would expose a deep 
and surprising role of supersymmetry in low energy hadron physics.  The prediction of orientifold equivalence for the 
size of the quark condensate in one-flavor QCD has been compared with lattice results in \cite{DHS}.  In our work, we 
consider another very interesting prediction of orientifold equivalence which, if valid, would be an unmistakable relic 
of the supersymmetry of the parent theory.  We refer to the prediction that the lowest-lying scalar and pseudoscalar 
mesons of one-flavor QCD (hereafter referred to as $\sigma$ and ${\eta}'$ respectively) should be degenerate.  In the 
parent SUSY theory, this degeneracy results from the simple fact that $\sigma$ and ${\eta}'$ belong to the same 
Wess-Zumino supermultiplet and that supersymmetry is unbroken.  This is conveniently described in terms of the 
supersymmetric chiral Lagrangian constructed by Veneziano and Yankielovich (VY) \cite{VY}.  In this formalism, the 
scalar and pseudoscalar mesons are associated with the real and imaginary parts of the complex scalar Wess-Zumino 
field representing gluino-antigluino bound states.  Although the scalar-pseudoscalar meson degeneracy is natural in 
the supersymmetric parent theory, it is much more mysterious from the point of view of one-flavor QCD.  In that theory, 
we are accustomed to thinking of the pseudoscalar meson ${\eta}'$ as a would-be Goldstone boson which acquires a mass 
via the axial anomaly.  In lattice calculations, the role of the anomaly in generating the ${\eta}'$ mass has been 
verified in detail by the study of the quark-line-disconnected "hairpin diagram" contributions to the pseudoscalar 
correlator.  In contrast to the would-be Goldstone boson interpretation of the ${\eta}'$, the scalar meson ${\sigma}$ 
is usually presumed to be in most respects a typical P-wave quark-antiquark meson.  Phenomenology and lattice calculations 
suggest that the corresponding flavor nonsinglet mesons are quite heavy, in the 1300-1500 MeV range.  The flavor singlet 
${\sigma}$ correlator includes hairpin diagrams along with the valence correlator which results in a large negative mass 
shift of the flavor singlet relative to the nonsinglet.  The scalar hairpin correlator was calculated in quenched 
lattice QCD in \cite{IT}  as part of a study of the spin-parity structure of the OZI rule.  Within the quenched 
approximation, \cite{IT} and \cite{BDEIT} contain all of the Monte Carlo results necessary for testing the orientifold 
prediction.  Although errors on the scalar hairpin mass shift are sizable, we show that within these errors, the 
combination of the upward shift of the pseudoscalar mass and the downward shift of the scalar mass, when evaluated 
in the one-flavor theory, brings the two mesons into approximate degeneracy. In order to try to explore the mechanism
responsible for the degeneracy, we also present the results of similar pseudoscalar and scalar meson flavor 
singlet mass calculations in two-dimensional $CP^3$.  
In this model it is relatively easy to use the overlap operator in high-statistics 
calculations to explore the chiral limit. 
This permits us to compare the scalar and pseudoscalar masses with much smaller statistical error.

\section{ORIENTIFOLD PLANAR EQUIVALENCE}

     The orientifold planar equivalence of SUSY Yang-Mills theory (SYM) and one-flavor QCD states that SYM is equivalent 
to one-flavor QCD to the extent that $N_c$ = 3 can be considered large.  This equivalence can most easily be seen 
by proceeding in two steps:

\begin{enumerate}
\item
Define the orientifold-A projection of SYM and show that this theory is precisely equal to one-flavor QCD for the special 
case of $N_c$ = 3.
\item
Demonstrate that for large $N_c$ the orientifold-A projection is equivalent to SUSY Yang-Mills.
\end{enumerate}

The orientifold-A projection can be understood by examining the SYM Lagrangian.

\begin{equation}
\label{eq:Lsym}
L_{SYM}=-\frac{1}{4g^2}G^a_{\mu\nu}G^a_{\mu\nu} + \frac{i}{g^2}\overline{\lambda}^a {\gamma}_{\mu} (D_{\mu}\lambda)^a
\end{equation}

Here  $\lambda^a = \left( \begin{array}{c} {\eta}^a_{\alpha} \\ \overline{\eta}^{\dot{\alpha} a} \end{array} \right)$ 
is the gluino, the fermionic super-partner of the gluon, written as a four-component Majorana spinor in the adjoint 
representation, and $D_{\mu}{\lambda}^a = \partial_{\mu}{\lambda}^a + gf^{abc}A_{\mu}^b{\lambda}^c$ is the adjoint 
covariant derivative.  Note that the SYM Lagrangian contains $(N^2-1)$ Weyl fermionic (gluino) degrees of freedom.  
The orientifold-A projection organizes these Weyl degrees of freedom in Dirac spinors associated with the 
$\frac{N(N-1)}{2}$ anti-symmetric generators of SU(N).

\begin{equation}
\label{eq:MtoD}
{\lambda}^i_j = {\lambda}^a(t^a)^i_j   \rightarrow   {\psi}^{[ij]} = {\psi}^{\tilde{a}}(t^{\tilde{a}})^{ij} = 
\left( \begin{array}{c} {\eta}^a_{\alpha} \\ \overline{\eta}^{\dot{\alpha} b} \end{array} \right)(t^{\tilde{a}})^{ij}
\end{equation}

Where:
\begin{itemize}
\item
$\tilde{a}$ is an index associated with the $\frac {N(N-1)}{2}$ anti-symmetric generators of SU(N)
\item
a,b are indices associated with the $(N^2-1)$ generators of SU(N)
\item
$(t^{\tilde{a}})^{ij}$ are the anti-symmetric generators of SU(N)
\end{itemize}

The gauge field and couplings are unchanged.   For $N_c$ = 3 there are three antisymmetric generators,
and the antisymmetric representation is the antitriplet. Thus, the 
orientifold-A projection is identical to one-flavor QCD.  Proof that the orientifold-A projection is equivalent 
to SYM for large $N_c$ is provided in \cite{ASV1,ASV2,ASV3}.

\section{THE EFFECT OF HAIRPIN DIAGRAMS}

The propagator for a flavor singlet meson in full QCD includes an arbitrary number 
of annihilation vertices.  In the quenched approximation all fermion lines must be attached to a creation/annihilation 
operator so only the valence and single-annihilation diagrams are present. If the coupling of fermion loops in the hairpin diagram is 
treated as a pure mass insertion $m_0^2$, then by geometric summation:

\begin{equation}
\label{eq:hpsum}
m^2_{\emph{full QCD}} = m^2_{\emph{valence}} \pm m^2_{\emph{hairpin}}
\end{equation}

The sign of the mass insertion is \emph{positive} for pseudoscalars because the operator 
$\overline{\psi}{\gamma}^5\psi$ is anti-hermitian and thus the valence correlator has a negative sign.  
The sign of the mass insertion is \emph{negative} for scalars since $\overline{\psi}\psi$ is hermitian.

\section{QCD MONTE CARLO RESULTS}

We have tested the prediction of scalar-pseudoscalar degeneracy in one-flavor QCD using one of the lightest quark masses 
from the studies \cite{IT,BDEIT}, corresponding to clover improved Wilson fermions with $C_{\emph{sw}} = 1.57$ with 
hopping parameter $\kappa = .1427$.  The calculation used 300 quenched gauge configurations generated with the Wilson 
action at $\beta = 5.7$ on a $12^3 \times 24$ lattice.  The modified quenched approximation (MQA) \cite{BDET} was used 
to resolve the problem of exceptional configurations.  Scalar and pseudoscalar hairpin diagrams were calculated using 
the "allsource" method \cite{KFMP}.  In this technique the quark propagator is calculated from a sum of identical unit 
color-spin sources located at all space-time points on the lattice.  When this propagator is contracted over color 
indices at a particular space-time point, the result is a gauge-invariant term corresponding to a closed quark loop 
originating at that point, plus a large number of gauge-dependent terms corresponding to open quark loops.  The 
gauge-dependent terms tend to cancel on average due to their random phases, permitting the calculation of closed quark 
loops and loop-loop correlators (hairpins).
   The results of these studies for $\kappa = .1427$ are presented in Table \ref{tab:QCDresults}.  The most striking 
finding is the large negative contribution of the hairpin to the mass of the scalar meson.  Although the statistical 
errors in the measurement of the scalar disconnected diagram are large, the size and sign of the scalar mass shift 
make scalar-pseudoscalar mass degeneracy plausible.  The analysis assumes the absence of excited states in both the 
scalar and pseudoscalar hairpin diagrams.  This assumption has been confirmed for the pseudoscalar in \cite{BDEIT}.

\begin{table}[h]
\centering
\caption{One-flavor QCD Results in $Mev^2$}
\begin{tabular}{||c|||c|c|c||} \hline \hline
$J^{PC}$  & ${Valence Mass}^2$     & ${Mass Shift}^2$    & ${Total Mass}^2$  \\ \hline \hline
$0^{-+}$ & $+[315(6)]^2$ & $+[407(11)]^2$ & $+[515+12-13]^2$ \\ \hline
$0^{++}$ & $+[1416(14)]^2$ & $-[1350(90)]^2$ & $+[427+249-756]^2$ \\ \hline
\end{tabular}
\label{tab:QCDresults}
\end{table}

\begin{figure}
\label{fig:plotQCD}
\includegraphics{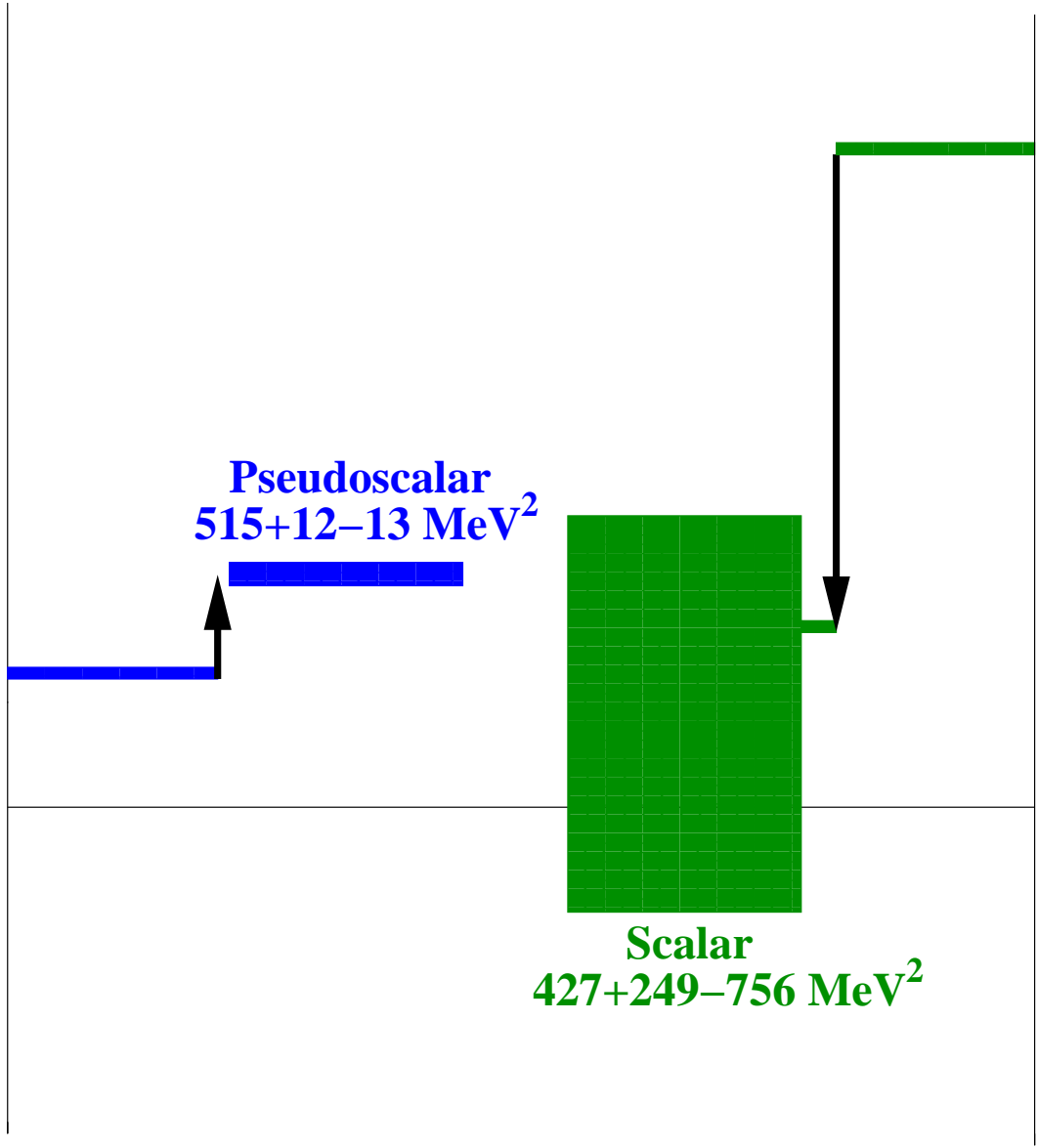}
\caption{Scalar and Pseudoscalar $Mass^2$ for One-Flavor QCD}
\end{figure}

\section{$CP^{(N-1)}$ MONTE CARLO RESULTS}

Although it is a 1+1 dimensional model, $CP^{(N-1)}$ shares several characteristics with QCD that make it an 
interesting testing ground for QCD concepts.  In addition to confinement and asymptotic freedom, $CP^{(N-1)}$ 
has a topological charge structure similar to QCD \cite{ALT}.  A major computational advantage is that we are 
able to use the overlap Dirac operator to calculate fermion propagators for light quarks with relative ease.  
$CP^{(N-1)}$ is a model of N complex scalar fields z subject to the constraint z*z=1.  The Lagrangian for 
$CP^{(N-1)}$ can be 
written in a form that includes a U(1) gauge field:  

\begin{equation}
\label{eq:CPNLag}
L = \beta (\partial_{\mu} - iA_{\mu})z^*_i(\partial_{\mu} + iA_{\mu})z_i
\end{equation}

\begin{equation}
\label{eq:CPNA}
A_{\mu} = \frac {i}{2} (z^*_i\partial_{\mu}z^i - z^i\partial_{\mu}z^*_i)
\end{equation}

This gauge field can be used to write a Lagrangian for $CP^{(N-1)}$ with fermions:
\begin{equation}
\label{eq:CPNFLag}
L = \overline{\psi^a}(i\gamma^{\mu}\partial_{\mu}-\gamma^{\mu}A_{\mu})\psi^a - m\overline{\psi^a}\psi^a = 
\overline{\psi^a}(i\gamma^{\mu}D_{\mu} - m)\psi^a
\end{equation}
Following the approach of the QCD studies, we treat the fermions in the quenched approximation, which is the proper
venue for studying hairpin vertices.

We calculated scalar and pseudoscalar correlators using 1000 gauge configurations on a $30 \times 30$ lattice 
with overlap fermions and $\beta = 1.0$.  By using smeared operators we were able to demonstrate that the hairpin 
diagrams did not contain excited states.  The pseudoscalar behavior is similar to QCD except that the mass measured
from the valence correlator approaches a non-zero constant in the chiral limit due to a quenched chiral loop
effect. As in QCD, fitting the quenched scalar valence correlator is complicated by the presence of an $\eta$'-$\pi$ 
intermediate state.  Much of the statistical error in the scalar mass can be traced to the 
removal of this quenched artifact.  The scalar is also heavier than the pseudoscalar, making it more difficult to
extract the mass accurately. A smeared operator analysis shows that the scalar hairpin correlator lacks excited states but that 
the vertex exhibits $p^2$ dependence.  Figure 2 
 is a plot of the quantities $m_p^2 + m_{0p}^2$ and $m_s^2 - m_{0s}^2$ in the range $M_{quark}$ = 0.03-0.10 along with 
a linear chiral extrapolation.

\begin{figure}
\label{fig:plotm2}
\begin{center}
\includegraphics[width=125mm]{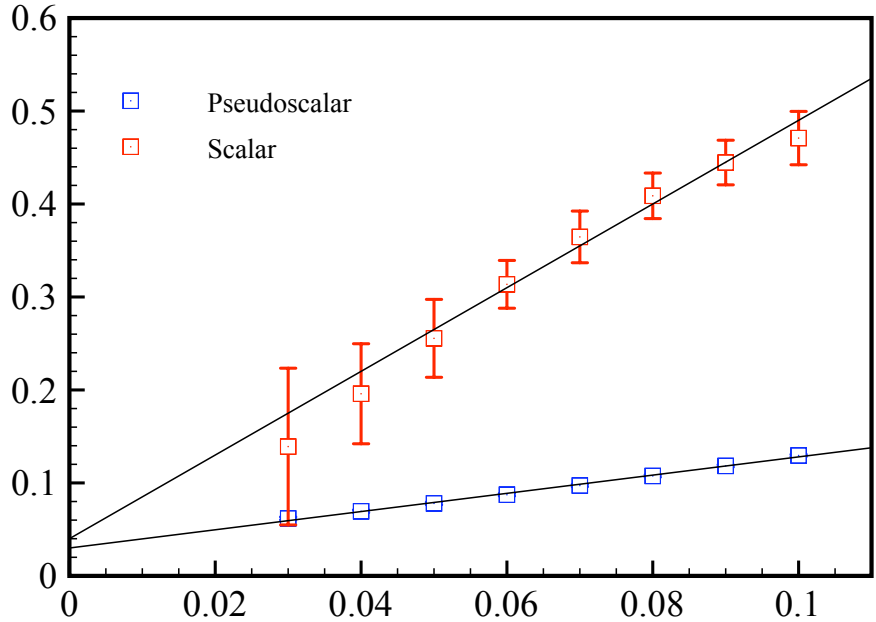}
\caption{$CP^{(N-1)}$ Meson $Mass^2$ vs Quark Mass}
\end{center}
\end{figure}

\section{DISCUSSION}

Although the statistical and systematic errors in our QCD calculations are significant, the important qualitative
result we have shown here is that the hairpin insertion diagram in the flavor singlet channel
corresponds to a positive mass shift for the pseudoscalar
meson and a negative mass shift for the scalar meson, and that the values obtained on the lattice for these
mass shifts are of roughly the right magnitude to bring an otherwise light pseudoscalar and an otherwise
heavy scalar into approximate degeneracy in 1-flavor QCD.
In the case of $CP^{(3)}$ our calculations also indicate approximate scalar-pseudoscalar degeneracy in the 
chiral limit.  The $CP^{(3)}$ data presented here include statistical errors only. The chiral extrapolation of the full 
scalar mass is clearly sensitive to the linear fit parameters.   Further study is required to quantify systematic errors 
in meson masses, particularly in the intermediate state subtraction in the scalar valence correlator.

\section{ACKNOWLEDGEMENTS}

This work was supported in part by the U.S. Department of Energy under grant DE-FG02-97ER41027.

\end{document}